\documentclass[vecphys]{svmult}
\usepackage{makeidx}         
\usepackage{graphicx}        
\usepackage{multicol}        
\usepackage[bottom]{footmisc}
\usepackage{natbib}
\makeindex             

\begin{document}
\title*{News from the ``Dentist's Chair'': Observations of AM\,1353-272 with the VIMOS IFU}
\titlerunning{``The Dentist's Chair''}
\author{Peter M.\ Weilbacher\inst{1} \and Pierre-Alain\ Duc\inst{2}}
\authorrunning{Weilbacher \& Duc}
\institute{Astrophysikalisches Institut Potsdam, An der Sternwarte 16, D-14482 Potsdam, Germany,
           \texttt{pweilbacher@aip.de}
      \and
           Service d'astrophysique, CEA Saclay-Orme des Merisiers, Bat.\ 709,
           91191 Gif sur Yvette cedex, France, 
           \texttt{paduc@cea.fr}
}
\maketitle

\begin{abstract}
The galaxy pair AM\,1353-272 nicknamed ``The Dentist's Chair'' shows two
$\sim$30\,kpc long tidal tails. Previous observations using multi-slit masks
showed that they host up to seven tidal dwarf galaxies. The kinematics of these
tidal dwarfs appeared to be decoupled from the surrounding tidal material. New
observations of the tip of the southern tidal tail with the VIMOS integral
field unit confirm the results for two of these genuine tidal dwarfs but raise
doubts whether the velocity gradient attributed to the outermost tidal dwarf
candidate is real.  We also discuss possible effects to explain the
observational difference of the strongest velocity gradient seen in the slit
data which is undetected in the new integral field data, but arrive at no firm
conclusion.
Additionally, low-resolution data covering most of the two interacting partners
show that the strongest line emitting regions of this system are the central
parts.
\end{abstract}

\section{Introduction}\label{PmW:Sect:Intro}
\begin{figure}
\centering
\includegraphics[width=9cm]{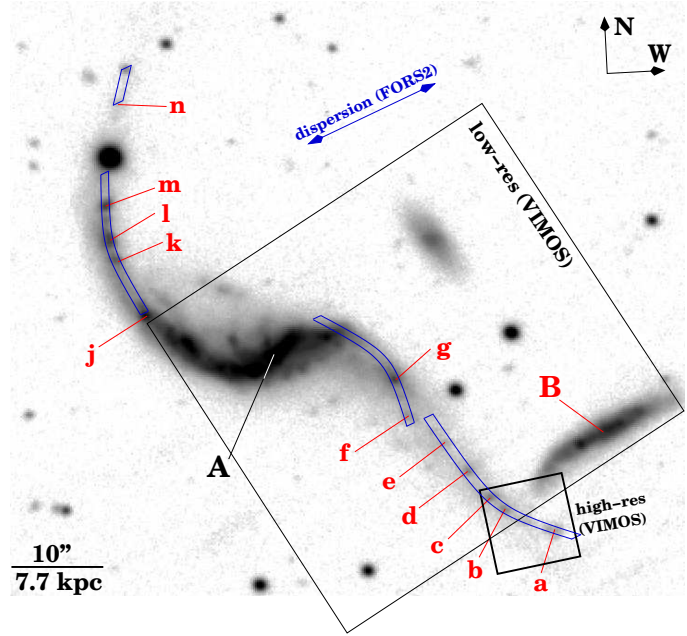}
\caption{The interacting system AM\,1353-272. The two interacting galaxies
(capital letters) and the relevant knots in the tidal tails are marked (lower
case letters). Overlayed are the original FORS2 slits and the two VIMOS
pointings in low and high resolution mode.}\label{PmW:Fig:System}
\end{figure}

Following old ideas about the creation of dwarf galaxies during interaction of
giant galaxies and detailed investigations of several nearby examples of these
Tidal Dwarf Galaxies \citep[TDGs,][]{DBS+00,DBW+97,DM98,HGvG+94}, we carried
out a first small survey of interacting galaxies \citep{WDF+00} with the aim of
better understanding the star-formation history of TDGs and constraining the
number of TDGs that are built per interaction \citep{WDF03}.  During this
survey, we studied a system cataloged as AM\,1353-272, which we called ``The
Dentist's Chair'' for its peculiar shape. Fig.~\ref{PmW:Fig:System} shows the
two components of the system, `A', a galaxy with $\sim$30\,kpc long tidal
tails, and `B', a disturbed disk galaxy, have a distance of $D \approx
160$\,Mpc ($H_0 = 75$\,km\,s$^{-1}$\,Mpc$^{-1}$). Within the tails several
obvious clumps with blue optical colors are visible. Using optical and
near-infrared imaging, evolutionary models, and optical spectroscopy,
\emph{seven} of these clumps were classified as TDG candidates in formation
\citep[][, marked `a' to `d' and `k' to `m']{WFDF02}. The largest velocity
gradient with an amplitude of $>$300\,km\,s$^{-1}$ appeared in TDG candidate
`a', at the very end of the southern tidal tail. This raised the question how
an object with relatively low luminosity could exhibit such fast ``rotation''.
However, as these observations were done using the multi-slit technique and
hence are spatially restricted due to the narrow slit, subsequent observations
were planned using an integral field unit (IFU) to cover more of the tidal
tails and view the velocity structure of the TDGs in two dimensions.

\section{IFU Observations}\label{PmW:Sect:Obs}
Our new VIMOS data taken at the ESO VLT consists of 1.3 hours of exposure time
in high resolution blue mode (field of view of
13\hbox{$^{\prime\prime}$}$\times$13\hbox{$^{\prime\prime}$}) and good seeing
conditions ($\sim$0\hbox{$.\!\!^{\prime\prime}$}7), targeted at the tip of the
southern tidal tail. This includes the TDG candidates `a' to `c'. The center of
galaxy `A', much of the southern tidal tail, and the companion `B' were
targeted with low-resolution (blue grism,
54\hbox{$^{\prime\prime}$}$\times$54\hbox{$^{\prime\prime}$}), and observed for
1 hour in mediocre conditions with $\sim$2\hbox{$.\!\!^{\prime\prime}$}0
seeing. These two pointings are sketched in Fig.~\ref{PmW:Fig:System}.

The data were reduced using the ESO pipeline for the VIMOS instrument. We made a
small enhancement to the code that allowed us to interpolate the wavelength
solution between adjacent spectra on the CCD. This was only used for the
low-resolution data, where the errors introduced with this method are smaller
than the accuracy allowed for by the spectral resolution. In low-resolution
mode, on the order of 15\% of the spectra could not be wavelength calibrated
due to overlapping spectral orders. From the final datacube of extracted and
wavelength calibrated spectra, we measured the relative fluxes and velocities
in each spectral element using Gaussian fits to the brightest usable emission
line. In the low-resolution data [O{\sc iii}]5007 is the strongest emission
line but at this redshift ($z\approx0.04$) it is strongly blended with a sky
emission line, so H$\beta$ had to be used instead. In the high-resolution data,
[O{\sc iii}]5007 is less affected by the sky-line and is the only line with
sufficient S/N for the analysis in the low surface brightness region near the
end of the tidal tail.

\section{Results}\label{PmW:Sect:Res}
\begin{figure}[t]
\centering
\includegraphics[height=4.5cm]{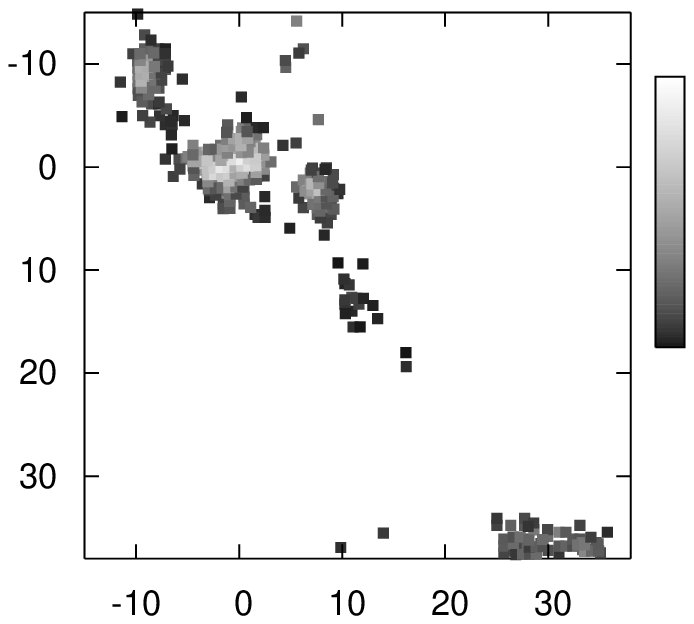}\hfill
\includegraphics[height=4.5cm]{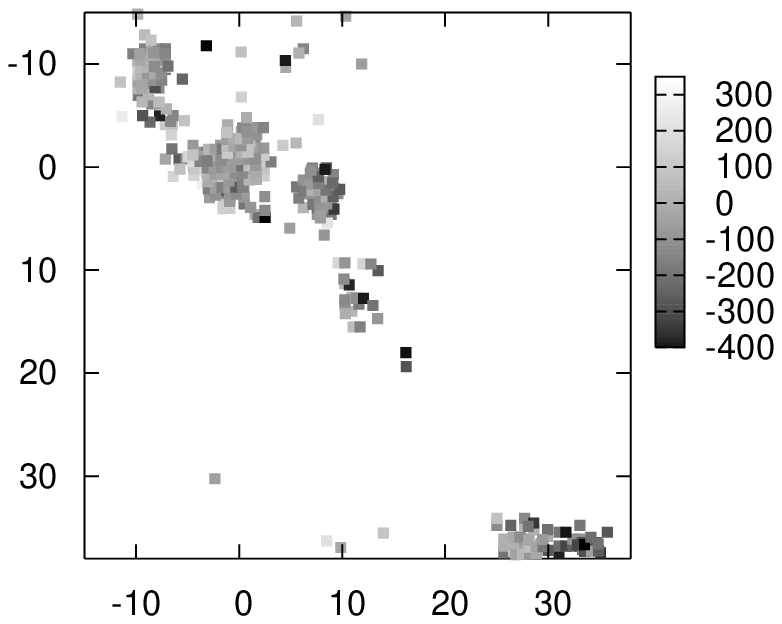}
\caption{Results of the low-resolution pointing, the center of AM\,1353-272\,A
is at the coordinates (0,0); at the bottom right, near (30,32) companion `B' is
visible. Axes labels are in arcseconds. {\bf Left}: relative H$\beta$ emission
line flux (bright: high flux, dark: low flux). {\bf Right}: velocity field
derived from H$\beta$ emission line (the greyscale bar gives relative
velocities in km\,s$^{-1}$).}\label{PmW:Fig:LowRes}
\end{figure}

Fig.~\ref{PmW:Fig:LowRes} summarizes the results that can be derived from the
low-resolution pointing using fits to the H$\beta$ emission line. The southern
tidal tail in undetected in this exposure and the strongest line emission
appears to be in the center of `A' and in the two knots at the end of its
bar-like central structure \citep[designated `g' and `j' in][]{WFDF02}. Galaxy
`B', despite being strongly reddened, also is a strong source of H$\beta$ line
emission. As the velocity resolution is on the order of 100\,km\,s$^{-1}$, in
this mode of VIMOS we cannot resolve the velocity structures in individual
knots, but the bar-like structure in `A' seems to rotate (the eastern end near
knot `j' is receding, the western end near knot `g' is approaching). The same
is true for the companion `B'.

\begin{figure}[t]
\centering
\includegraphics[width=7cm,angle=270]{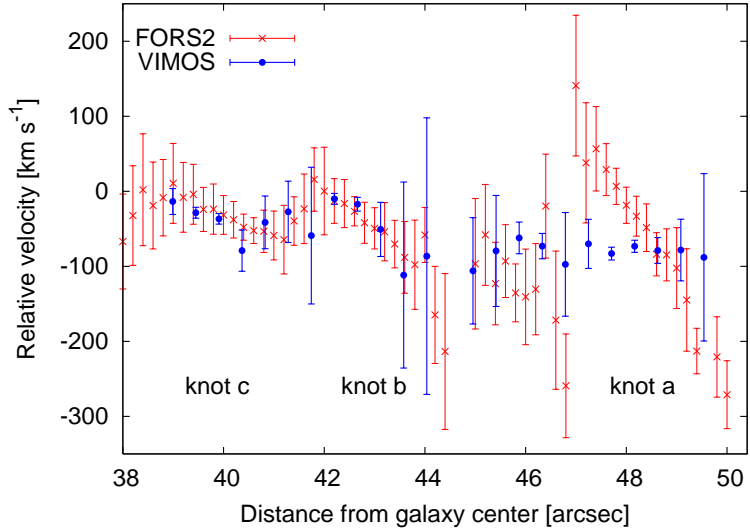}
\caption{Velocity field as derived from FORS2 observations (black points) and
reconstruction of this velocity field from the VIMOS IFU datacube (grey
crosses).}\label{PmW:Fig:OIIIvelo}
\end{figure}

To verify our original FORS2 multi-slit observations, we try to ``reconstruct''
them from the VIMOS datacube. To that end we average over the spaxels
approximately in the dispersion direction of the FORS2 observations and derive
an average redshift over the slit width.  Fig.~\ref{PmW:Fig:OIIIvelo} shows the
resulting velocity field along this artificial slit and compares it with the
velocity profile of the FORS2 data. From this plot it can be seen that the
velocity gradients of knots `b' and `c' are very well recovered, while the
steep slope within knot `a' with an amplitude of $\sim$350\,km\,s$^{-1}$ in the
FORS2 data appears almost flat in the reconstructed VIMOS data.  We tested
several alternative slit positions and curvatures, varied the effective slit
width, and also tried to add slit effects (velocity offsets due to non-centered
emission within the slit) to our reconstruction. None of these changes improved
the match for knot `a'. In fact, slit effects significantly worsened the
agreement for all three knots. The FORS2 observations were done on these
extended objects in 1\hbox{$.\!\!^{\prime\prime}$}0 seeing with a
1\hbox{$.\!\!^{\prime\prime}$}2 slit, so that it appears unlikely that slit
effects would have a strong contribution to the observed velocity gradient.
Other problems, like instrumental flexures should have been removed by the data
reduction procedure as detailed in \citet{WFDF02}. We are therefore confident
that slit effects do not play an important role in the FORS2 data. On the other
hand, if we assume that the original slit-based data give the correct results
it is unclear, how the VIMOS data could be flawed to hide this one velocity
gradient. The wavelength calibration works well for the high-resolution mode as
confirmed by checks with sky emission lines.

\section{Summary \& Outlook}\label{PmW:Sect:Sum}
We presented a few tentative results for the interacting system called ``The
Dentist's Chair'' from new observations with the VIMOS integral field mode:
line emission seems to be                          
concentrated within the centers of the interacting partners while the tidal
tails themselves are not detected H$\beta$ narrowband slices.
Three knots, previously identified as TDG candidates near the end of the
southern tidal tail, are detected in [O{\sc iii}]5007 emission. For two of them
the velocity profiles were confirmed.
However, the strongest velocity gradient in the outermost TDG candidate (knot
`a') as measured on FORS2 data is not confirmed by the VIMOS datacube. The
reason for this discrepancy is unknown.

To solve this mystery and find more clues to the origin of the velocity fields
seen in this interacting system, further, deeper IFU observations, taken with
appropriate dither offsets to facilitate more accurate sky subtraction, are
required. With other instruments like e.\,g.\ the GMOS IFU would be possible to
cover both the Ca-triplet and H$\alpha$ in the same exposures and directly
compare the stellar velocity field with ionized gas dynamics. As good S/N is
required to detect the absorption lines, this can only be done in the brighter
northern tidal tail.

\begin{acknowledgement}
PMW received financial support through the D3Dnet project from the German
Verbundforschung of BMBF (grant 05AV5BAA). We are grateful to Ana Monreal-Ibero
and Lise Christensen for practical hints on IFU data handling. The data we
discuss was taken in service mode at Paranal (ESO Program 074.B-0629).
\end{acknowledgement}


\bibliographystyle{../../AA}
\bibliography{../../PmW}


\printindex
\end{document}